\documentclass[final]{ias2}

\usepackage{graphicx} 
\usepackage{multirow}
\usepackage{array} 

\newcommand{\be}{\begin{equation}}
\newcommand{\ee}{\end{equation}}
\newcommand{\bea}{\begin{eqnarray}}
\newcommand{\eea}{\end{eqnarray}}
\newcommand{\bra}{\langle}
\newcommand{\ket}{\rangle}
\newcommand{\hm}{\hspace*{-0.2cm}}
\newcommand{\nn}{\nonumber}
\newcommand{\re}{{\rm Re}}
\newcommand{\im}{{\rm Im}}

\usepackage{hyperref} 

\begin{document}

\markboth{Lattice QCD at high temperature and density}{Gert Aarts}

\title{
Developments in lattice QCD  for matter 
\\ 
\vspace*{-0.2cm}
\\
 at high temperature and density}

\author[ga]{Gert Aarts} 
\email{g.aarts@swan.ac.uk}
\address[ga]{Physics Department, College of Science, Swansea University, Swansea SA2 8PP, United Kingdom}

\begin{abstract}
A brief overview of the QCD phase diagram at nonzero temperature and density is provided. It is explained why standard lattice QCD techniques are not immediately applicable for its determination, due to the sign problem.
We then discuss a selection of recent lattice approaches that attempt to evade the sign problem and classify them according to the underlying principle:
constrained simulations (density of states, histograms), 
holomorphicity (complex Langevin, Lefschetz thimbles),
partial summations (clusters, subsets, bags)
and
change in integration order (strong coupling, dual formulations).

\end{abstract}


 
\maketitle


\section{Introduction}

Establishing the QCD phase diagram at nonzero temperature and density remains one of the outstanding challenges in the theory of the strong interactions. As is well known \cite{Kogut:2004su,Yagi:2005yb}, its structure is relevant for heavy-ion collisions and the creation of the quark-gluon plasma at the Relativistic Heavy Ion Collider (RHIC) at Brookhaven, the Large Hadron Collider (LHC) at CERN, and the forthcoming  Facility for Antiproton and Ion Research (FAIR) at GSI, for neutron stars and compact objects, and for the early Universe.
From a theoretical perspective, the main obstacle in its determination is that numerical lattice QCD, the prime nonperturbative tool to study QCD, cannot be used in conjunction with standard numerical techniques. Lattice QCD relies on importance sampling, assigning a real and positive number  to each configuration of quarks and gluons.  
At nonzero (and zero) temperature but vanishing chemical potential, this approach is indeed possible, since the Boltzmann weight in the QCD partition function is real and positive. The result is a detailed understanding of the crossover between the hadronic and quark-gluon plasma phase as the temperature is increased \cite{Aoki:2006we,Borsanyi:2010cj,Bazavov:2011nk,Philipsen:2012nu}. In recent years, the phase structure has also been studied in the presence of an external magnetic field, relevant for cosmology and noncentral heavy-ion collisions \cite{Kharzeev:2012ph}.
Since the Boltzmann weight remains real and positive, lattice techniques can be applied here as well 
\cite{D'Elia:2010nq,Bali:2011qj}.
 
However, as soon as the baryon chemical potential is nonzero, the quark contribution to the Boltzmann weight  is complex (in the usual formulation), ruling out conventional approaches. This issue is usually referred to as the sign problem or complex-action problem. It is certainly a nontrivial problem, since an incorrect handling of the sign problem results in manifest failure and physically wrong results. This realisation has been around since the first lattice simulations at nonzero chemical potential were carried out, in the 1980s \cite{Barbour:1986jf}.  It is therefore also a persistent problem, for which no definite solution has been found yet.
Interestingly, the sign problem is not specific to QCD but appears in many lattice theories with a mismatch in particle densities, which makes it relevant from a more broader perspective as well.

In this overview, I will first remind the reader of some general observations regarding the QCD phase diagram and the sign problem. Subsequently I will discuss various approaches that have been pursued in recent years and attempt to categorise them according to the underlying principle. 
Refs.\ \cite{deForcrand:2010ys,Aarts:2013bla} contain partially complementary reviews of the sign problem in QCD at  finite density.

\section{QCD phase diagram and the sign problem}

\begin{figure}[t]
 \begin{center}
  \includegraphics[height=6cm]{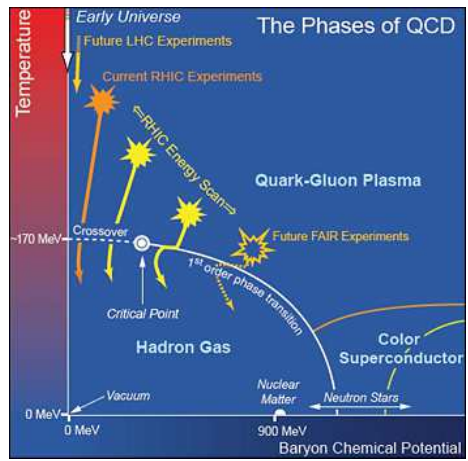}
 \caption{Impression of the QCD phase diagram \cite{fig}.}
\label{fig:pd}
\end{center}
\end{figure}

An impression of the QCD phase diagram is given in Fig.\ \ref{fig:pd}. Along the vertical temperature ($T$) axis, the crossover between the hadronic  phase and the quark-gluon plasma can be seen. Recent lattice studies agree that the crossover temperature is closer to 155 MeV than 170 MeV, as indicated in the figure 
\cite{Borsanyi:2010cj,Bazavov:2011nk}. Increasing the chemical potential, the thermal crossover may change in a proper first-order phase transition, with a second-order critical endpoint to mark where this happens   \cite{Stephanov:2004wx}.
Along the horizontal chemical potential ($\mu$) axis, transitions to new phases are expected: nuclear matter, various colour superconducting phases   \cite{Alford:2007xm} and a possible quarkyonic phase  \cite{McLerran:2007qj}  (not indicated). Phase transition lines can join in a triple point. Heavy-ion collider experiments probe (or will probe) various regions of the phase diagram, depending on the energy of the collisions.

However, due to the sign problem most features of the phase diagram have not been obtained using lattice methods, but instead follow from general considerations and predictions from effective models (such as Polyakov-loop extended Nambu Jona-Lasinio (PNJL)  \cite{Fukushima:2003fw,Ratti:2005jh} and Polyakov-quark-meson models \cite{Herbst:2010rf}), the functional renormalization group \cite{Braun:2009gm}, Schwinger-Dyson equations \cite{Fischer:2012vc},
or a combination therefore \cite{Fischer:2013eca}.
They have therefore not been established unequivocally.
Of course, at asymptotically high energy scales, asymptotic freedom supports the validity of weak-coupling computations \cite{Vuorinen:2003fs,Kurkela:2009gj,Andersen:2011sf}.

The sign problem appears in the standard lattice formulation of QCD, in which the partition function is written as
\be
\label{eq:Z}
Z = \int DU D\bar\psi D\psi \, e^{-S_{\rm YM} - S_{\rm F}}  = \int DU\,  e^{-S_{\rm YM}} \det M.
\ee
Here the gauge links $U$ represent the gluonic degrees of freedom, with the Yang-Mills action $S_{\rm YM}$, and $\psi, \bar\psi$ are the quark fields, with the fermionic action $S_{\rm F} \sim \bar\psi M(U;\mu)\psi$. 
The chemical potential appears in the fermion matrix $M$.
In the second expression, the quark fields have been integrated out exactly, resulting in the quark determinant.  It follows from elementary properties of the Dirac operator that the determinant is complex when $\mu$ is real and nonzero,  and satisfies \cite{deForcrand:2010ys}
\be
\label{eq:det}
[\det M(\mu)]^* = \det M(-\mu^*),
\quad\quad\quad
 \det M(\mu) =  |\det M(\mu)| e^{i\vartheta},
\ee
i.e.\ it has a complex phase. It can be seen that the determinant is real when the chemical potential is purely imaginary, such that standard simulations are possible in this case \cite{Lombardo:1999cz}.

Lattice QCD has been effective in the crossover region at nonzero temperature  for small chemical potentials, relying on analyticity in $(\mu/T)^2$. In this approach, simulations are carried out at either vanishing or imaginary chemical potential, where the sign problem is absent. Applying a Taylor expansion in $(\mu/T)^2$ requires the calculation of higher-order derivatives of the logarithm of the fermion determinant with respect to $\mu$, i.e.\ of generalised susceptibilities \cite{Gupta:2011ma}. 
Even though these quickly become very noisy, the expansion coefficients have been
determined up to the impressive order of $(\mu/T)^8$, and from these the position of the critical endpoint has been predicted, based on an estimate of the radius of convergence of the Taylor expansion \cite{Datta:2012pj}.
This prediction should be contrasted with the conclusion from studies at imaginary chemical potential  \cite{deForcrand:2006pv} and lower-order Taylor expansions \cite{Endrodi:2011gv}, 
where it is found that the crossover weakens as $(\mu/T)^2$ is increased from zero, making it less likely that the critical point can be found in this way.

The interesting region in the phase diagram with regard to the sign problem is the region at lower and zero temperature. The reason is that in this region it is very well understood what to expect and also what will go wrong when the sign problem is not resolved correctly \cite{Stephanov:1996ki}.
 The onset to nuclear matter at vanishing temperature takes place when the baryon chemical potential equals the nucleon mass (minus the binding energy). For chemical potentials below this onset value, thermodynamic quantities should be independent of chemical potential (at zero temperature). In particular the baryon density has to vanish. The reason is simple: there is not enough energy available to create a baryon. At nonzero temperature, the baryon density is nonzero but Boltzmann suppressed, $\sim \exp[-(m_B-\mu)/T]$, where $m_B$ is a generic baryon mass.
When the sign problem is not correctly taken into account in lattice QCD simulations, this expectation is easily upset. For instance, when the complex phase of the determinant is ignored (phase quenching), one finds, in the case of two quark flavours with equal mass, that the onset takes place at a baryon chemical potential of 3/2 times the pion mass, rather than around the nucleon mass. Since the former is much less than the latter, one observes an early incorrect onset. This feature is well understood: ignoring the complex phase  results in a theory not at nonzero baryon chemical potential, but instead at nonzero isospin chemical potential, which couples to the pion rather than the nucleon \cite{Son:2000xc}. It implies that the inclusion of the complex phase of the determinant is necessary for the exact cancellation of this early onset to take place \cite{Osborn:2005ss,Splittorff:2007ck}. How this happens in lattice QCD simulations is not immediately obvious. It has therefore been dubbed the Silver Blaze problem \cite{Cohen:2003kd}.
Its correct resolution provides a stringent test of the validity of the applied approach \cite{Aarts:2013bla}.

Finally, we mention that the sign problem is regarded as severe when the expectation value of the phase factor in the phase-quenched (pq) theory goes to zero exponentially in the thermodynamic limit. This quantity is given by the ratio of the  partition functions with and without the complex phase \cite{deForcrand:2010ys,Aarts:2013bla},
\be
\bra e^{i\vartheta}\ket_{\rm pq}  
= \frac{\int DU\,  e^{-S_{\rm YM}} |\det M| e^{i\vartheta}}{ \int DU\,  e^{-S_{\rm YM}} |\det M|}
= \frac{Z}{Z_{\rm pq}} 
=  e^{ -V\Delta f/T},
\ee
 where $\Delta f$ is the difference in free energy densities between the original and the phase-quenched theories and $V$ is the three-volume. From the extensivity of free energies, this exponential behaviour is unavoidable.
 Furthermore, since $\Delta f$ can be reliably computed in chiral perturbation theory, the volume and  temperature dependence is understood precisely \cite{Splittorff:2006fu,Splittorff:2007zh}.
 It follows that approaches based on ignoring the phase are guaranteed to fail for large volumes and/or low temperature. This is for instance the case for reweighting methods \cite{Barbour:1997ej,Fodor:2001au}, in which part of the complex weight, such as the phase of the determinant, is absorbed in the observable, 
\be
\bra O\ket = \frac{\int DU\,  e^{-S_{\rm YM}} |\det M| e^{i\vartheta} O}{ \int DU\,  e^{-S_{\rm YM}} |\det M| e^{i\vartheta}}
= \frac{\bra e^{i\vartheta}O\ket_{\rm pq} }{\bra e^{i\vartheta}\ket_{\rm pq} }.
\ee
 In the presence of a Silver Blaze problem, the sign problem will certainly be severe, since excessive cancellations are required \cite{Splittorff:2007ck}, see also the review \cite{Splittorff:2006vj}.

In the following I discuss various approaches that have been proposed as complete or partial solutions to the sign problem in recent years and are currently under investigation. Since the sign problem is not unique for QCD, often these approaches are first tested in simpler models, for instance lower-dimensional ones. In particular, the sign problem may also appear in bosonic theories with a nonzero chemical potential and hence the presence of fermionic degrees of freedom is not essential.  I will only discuss a selection of approaches, referring again to Refs.\ \cite{deForcrand:2010ys,Aarts:2013bla} for complementary overviews.

\section{Density of states, histograms}

We start with a set of approaches which stay as close as possible to the original lattice formulation (\ref{eq:Z}), and which go under the headings of density of states,  factorisation or histogram method
\cite{Fodor:2007vv,Anagnostopoulos:2010ux,Ejiri:2013lia}.
 The idea is to evaluate the path integral in two stages, namely by first evaluating a constrained integral where one degree of freedom is fixed and subsequently performing the remaining integral over the resulting probability distribution -- the density of states -- constructed in the first step. The density of the states can be obtained by constructing histograms during the constrained simulation.  Explicitly, if we take as degree of freedom a generic observable $P$, such as the plaquette, action density or  Polyakov loop, the (unnormalised) density of states is given by
\be
\label{eq:w}
w(P) =  \int DU\, \delta(P-P') \, e^{-S_{\rm YM}} \det M,
\ee
where $P'$ is the value of $P$ taken during the simulation. The expectation value of $P$ is then determined by the simple integrals
\be
\bra P\ket = \frac{1}{Z} \int dP\, w(P) P,
\quad\quad\quad\quad
Z= \int dP\, w(P).
\ee
Variations of this approach have been discussed several times in the past \cite{Fodor:2007vv,Anagnostopoulos:2010ux}. A recent review on the histogram method can be found in Ref.\ \cite{Ejiri:2013lia}.
 The difficulty is that it is still needed to evaluate the constrained integral (\ref{eq:w}) with a complex weight at nonzero chemical potential. On small volumes, this may be handled by reweighting \cite{Fodor:2007vv}. Another possibility is to combine the approach with the Taylor expansion of the determinant in $(\mu/T)^2$ \cite{Ejiri:2007ga}.

It is also possible to use as observable the angle of the phase factor of the determinant itself \cite{Ejiri:2007ga}. We can then write
\bea
w(\vartheta) =&&\hm  \int DU\, \delta(\vartheta-\vartheta') \, e^{-S_{\rm YM}} \det M \nn\\
=&&\hm \int DU\, \delta(\vartheta-\vartheta') \, e^{-S_{\rm YM}} |\det M |  e^{i\vartheta'}  = e^{i\vartheta} w_{\rm pq}(\vartheta).
\eea
 The advantage is now that the density of states for the angle can be determined in the theory without the complex phase, i.e.\ in the phase-quenched theory. In Refs.\ \cite{Ejiri:2007ga,Saito:2013vja}
this approach is pursued using a cumulant expansion, focussing on the assumption that the phase-quenched density $w_{\rm pq}(\vartheta)$ is a simple Gaussian, so that only the lowest-order cumulant, the variance $\bra \vartheta^2\ket_c$, has to be determined. However, this assumption has been critically assessed recently: 
Ref.\ \cite{Greensite:2013gya} finds that nongaussian corrections are present in general. Even though these corrections are suppressed by powers of the volume in $w(\vartheta)$, they contribute at leading order to the final result obtained by integration over $\vartheta$. Therefore it is mandatory that the density of states is determined with great precision over a wide range \cite{Greensite:2013gya}. Here improvements to the original Wang-Landau algorithm \cite{WL} might be of use \cite{Langfeld:2012ah} and it has been conjectured that those should be applied to theories with a severe sign problem \cite{Langfeld:2013xbf}.

\section{Lefschetz thimbles, complex Langevin}

The partition function  (\ref{eq:Z}) is a path integral with a complex integrand. Simpler complex integrals are often evaluated by methods of steepest descent or by stationary-phase approximations, with saddle points situated in the complex plane rather than on the original real manifold, using holomorphic properties of the integrand and the ensuing freedom to change the integration contour. This suggests that it makes sense to view the path integral more generally and allow the degrees of freedom to become complex-valued. For QCD this implies that the gauge links $U$ are no longer limited to be in SU(3), but can now take value in the complex extension SL(3, $\mathbb{C})$. 
This is easily achieved. Writing $U = e^{i\lambda_a A_a}$,
 with $\lambda_a$ the Gell-mann matrices, $A_a$ is real-valued for SU(3), but complex-valued for  SL(3, $\mathbb{C})$. One consequence is that the links are no longer unitary, but instead satisfy Tr $UU^\dagger/3\geq 1$  \cite{Aarts:2008rr}.
In this approach, the partition function therefore keeps its original form, but the paths of integration are allowed to be more general.

This idea has been pursued in the past few years using two different approaches: stochastic quantisation (or complex Langevin dynamics) \cite{Parisi:1984cs,Klauder:1983}
and integration along Lefschetz thimbles \cite{Cristoforetti:2012su}.
 I will first discuss the second approach, which closely follows the method of steepest descent and has a long history in complex integrals, such as the Airy function \cite{Witten:2010cx}.
The goal is to deform the path of integration, passing through the classical saddle points, such that the imaginary part of the action (which includes the logarithm of the quark determinant in the case of QCD) is constant. 
The union of all the paths passing through a saddle point make up a thimble ${\cal J}$. Since the imaginary part of the action is constant along the thimble, it can be taken out of the integral.
For one degree of freedom and one saddle point, this amounts to writing
\cite{Cristoforetti:2012su,Mukherjee:2013aga}
\bea
Z =&&\hm \int dx\, e^{-S(x)} = e^{-i\im S_{\cal J}} \int_{\cal J} dz\, e^{-\re S(z)}
\nn\\
=&&\hm  e^{-i\im S_{\cal J}} \int ds\, J(s) e^{-\re S(z(s))},
\quad\quad\quad J(s) = x'(s)+iy'(s).
\eea
 In the second line the thimble is parameterised explicitly in terms of $z(s)=x(s)+iy(s)$.  Due to the curvature of the thimble, there is a complex jacobian $J(s)$, leading to a residual sign problem, but this may be milder than the original one  \cite{Cristoforetti:2012su}. Clearly more dangerous is the situation where more than one saddle point and associated thimble contributes: in that case there are  relative phase differences between the contributions from the thimbles, yielding again a possible severe sign problem.
 Based on universality, it has been conjectured that a single saddle point (e.g.\ the perturbative one) suffices \cite{Cristoforetti:2012su}.
Besides in toy models, so far the method has been tested in four-dimensional scalar field theory at nonzero chemical potential \cite{Cristoforetti:2013wha,Fujii:2013sra},
 where agreement with previous results \cite{Aarts:2008wh,Gattringer:2012df} has been found.

\begin{figure}[t]
 \begin{center}
  \includegraphics[height=4.4cm]{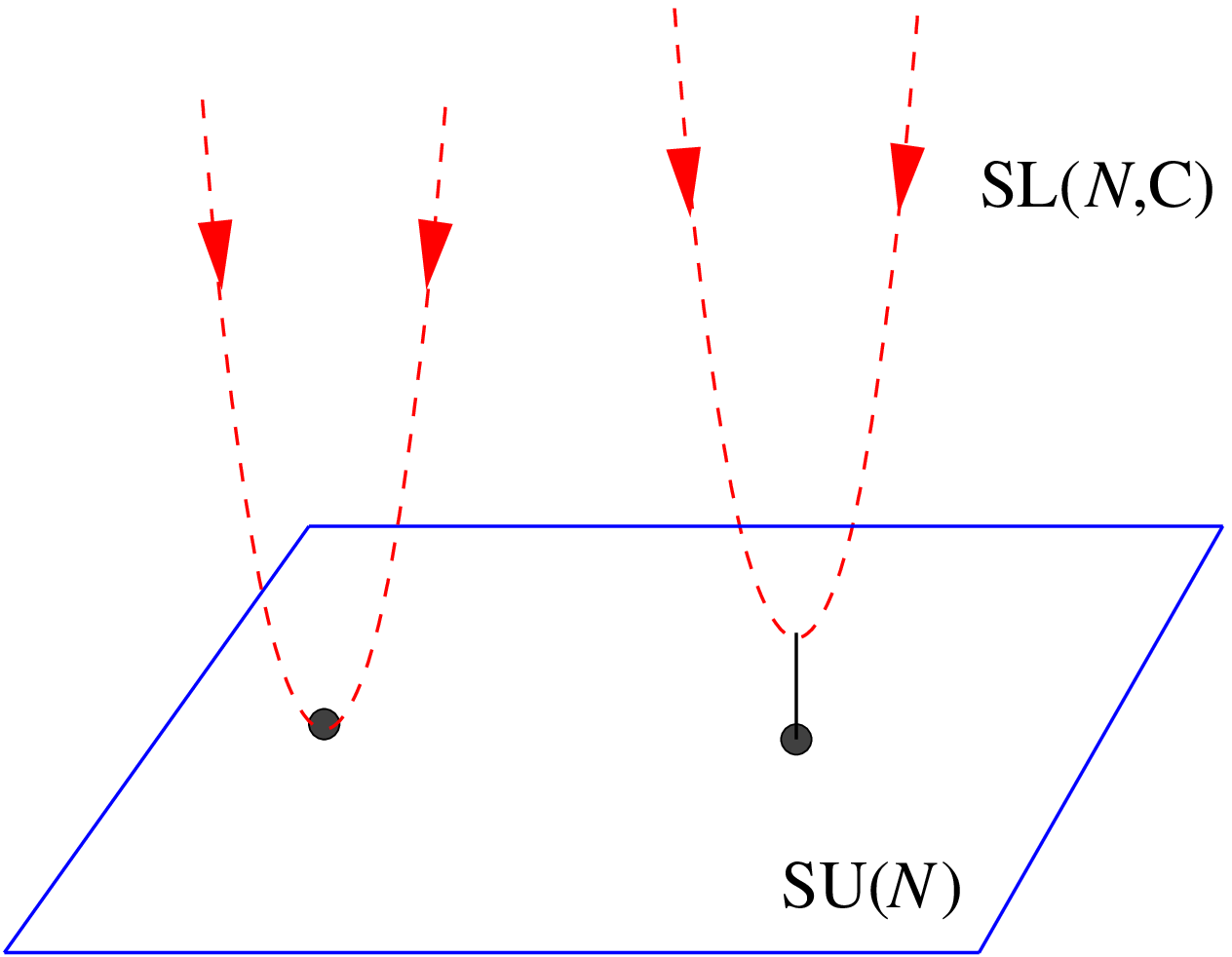}
  \includegraphics[height=4.4cm]{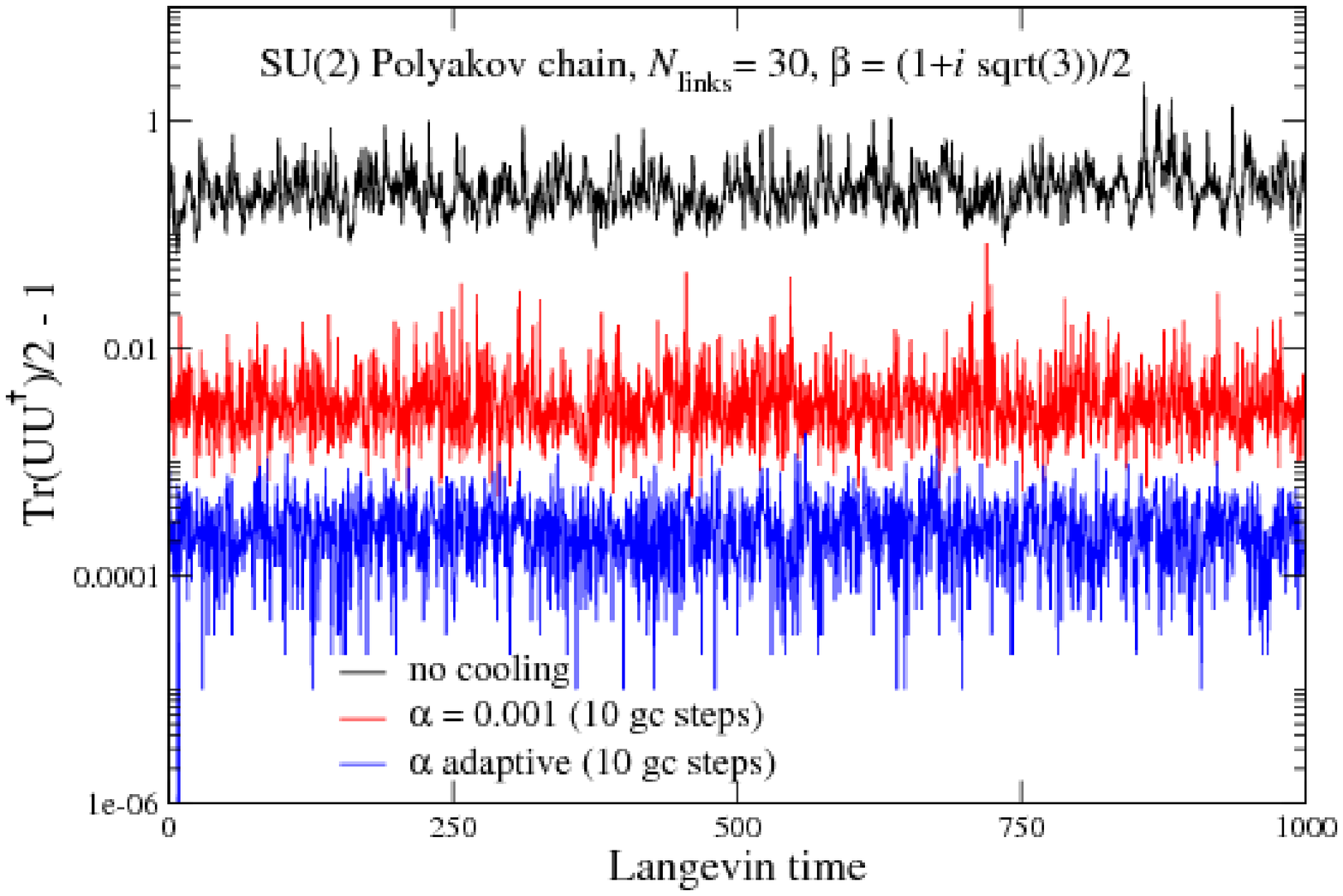}
 \caption{Gauge cooling in complex Langevin dynamics brings the links closer to the unitary submanifold (left), as demonstrated in an effective SU(2) model (right) \cite{Aarts:2013uxa}.}
\label{fig:gc}
\end{center}
\end{figure}

In stochastic quantisation/complex Langevin dynamics  \cite{Parisi:1984cs,Klauder:1983}, the complexified manifold is explored in a different manner, namely by using a  stochastic  process with complex drift terms, derived from the complex action. During this process, a real and positive probability distribution is effectively sampled.  
For one degree of freedom with a complex action, this means
\be
 Z = \int dx\, e^{-S(x)} \to \int dxdy\, P(x,y), 
 \quad\quad\quad
 P(x,y) \geq 0.
 \ee
One problem is that this distribution $P(x,y)$ is not known a priori, but is constructed during the stochastic evolution. However, the most important hurdle is that  it is not guaranteed that the process will converge to the physically correct result and incorrect convergence has indeed been observed \cite{Ambjorn:1986fz,arXiv:1005.3468}. During the past years, this conundrum has been clarified: it is now understood that correct results are obtained provided the distribution is well localised in the extended manifold and drops sufficiently fast at large distances from the real manifold \cite{arXiv:0912.3360,Aarts:2011ax}.
This insight has been used in a constructive manner for gauge theories, where SL($N, \mathbb{C}$) gauge cooling has been proposed as a tool to control the distribution sampled during the process \cite{Seiler:2012wz}. This idea is sketched in Fig.\ \ref{fig:gc} (left) and demonstrated in Fig.\ \ref{fig:gc} (right) for an effective SU(2) model with a complex coupling, where it is shown that gauge cooling, possibly adaptive   \cite{Aarts:2013uxa}, controls the distance from the unitary submanifold, for which Tr $UU^\dagger/2=1$ in this case. 
More details on complex Langevin dynamics can be found in the reviews  \cite{Aarts:2013bla,Aarts:2013uxa}.

Complex Langevin dynamics with gauge cooling has been applied successfully to QCD in the presence of static quarks at nonzero density \cite{Seiler:2012wz}.
Recently, first results for QCD with two and four flavours of light quarks have appeared as well \cite{Sexty:2013ica}. This is a major step forward, since it is the first simulation with light quarks at finite gauge coupling directly at nonzero density. The method successfully describes the transition from zero density all the way to saturation (the maximal density on the lattice). Very recently, preliminary results for SU(3) Yang-Mills theory in the presence of a $\theta$ term have also appeared \cite{Bongiovanni:2013nxa}
 and agreement with expected results from imaginary $\theta$, where the sign problem is absent, has been found. These results strongly suggest that complex Langevin dynamics with gauge cooling deserves to be studied more intensely.

There are a  number of open questions. Gauge cooling appears to be effective only when the gauge coupling is not too small, i.e.\ on fine lattices \cite{Seiler:2012wz}. This is not a problem in principle, but large lattice volumes are then required to avoid finite-size effects. It would also be useful to understand the origin of this finding. Furthermore, the presence of the logarithm of the determinant spoils the holomorphicity of the action. It has been argued that this results in an ambiguity for small quark masses, when the branch cut of the logarithm is crossed frequently  \cite{Mollgaard:2013qra}. This certainly needs to be understood better.

\begin{figure}[t]
 \begin{center}
  \includegraphics[height=4.cm]{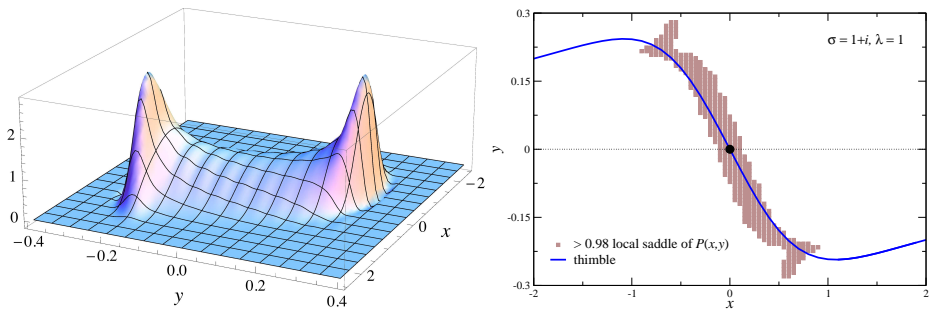}
  \includegraphics[height=4.cm]{plot-thimble-CL1-v3.eps}
 \caption{Distribution $P(x,y)$ effectively sampled during a complex Langevin process for the model (\ref{eq:quartic}) \cite{Aarts:2013uza} (left) and a comparison between the local saddle of this distribution and the thimble passing through the saddle point at the origin (right)  \cite{Aarts:2013fpa}.
 }
\label{fig:comp}
\end{center}
\end{figure}

Complex Langevin dynamics and the Lefschetz thimble approach both explore a larger complexified manifold to evade the sign problem. It is therefore interesting to compare the two approaches. Recently this has been done for a simple complex integral, where most results can be obtained analytically \cite{Aarts:2013fpa}. The partition function is
\be
\label{eq:quartic}
Z = \int_{-\infty}^\infty dx\, e^{-S(x)},
\quad\quad\quad\quad
S(x) = \frac{\sigma}{2}x^2 +\frac{\lambda}{4}x^4,
\ee
with a complex mass parameter $\sigma$. In Fig.\ \ref{fig:comp} (left) the distribution $P(x,y)$ sampled during the Langevin process is shown \cite{Aarts:2013uza}. It is strictly zero  when $y$ is larger than 0.3029 (for this choice of parameters), and based on the mathematical justification \cite{arXiv:0912.3360,Aarts:2011ax},  the method is then reliable. The thimble passing through the saddle point at the origin is shown in Fig.\ \ref{fig:comp}  (right). Also indicated is the region where $P(x,y)$ is largest. It is clear that the distribution and the thimble follow each other closely \cite{Aarts:2013fpa}. One may therefore wonder how generic this is and whether this observation can be used constructively.

\section{Clusters, subsets, fermion bags}

Let us go back again to the partition function (\ref{eq:Z}). The reason importance sampling is not applicable 
is that configurations come with a complex weight. However, it might be possible that for certain carefully selected sets of configurations, the combined weight is real and positive. The challenge then is to identify those sets (or clusters, subsets, bags, $\ldots$) and perform the path integral in two stages: first analytically construct the sets, and then numerically integrate over those using importance sampling, which is now possible since the combined weights are real and positive. This is the basic idea, successfully applied in the meron-cluster algorithm some time ago \cite{Chandrasekharan:1999cm}.

In recent years, this notion has been pursued in various forms. Here I will discuss two implementations. In the gauge theory without quarks, an important role is played by the $\mathbb{Z}_3$ centre symmetry, a global symmetry which is spontaneously broken at high temperature. 
Both the gauge action and the Haar measure in the path integral are invariant under centre transformations, $U\to e^{2\pi i k/3}U$, $k=0,1,2$. One can therefore trivially form a subset  identifying the configurations related by centre symmetry.
 This gets more interesting in the presence of quarks, since the quark contribution is not invariant under the symmetry and hence the configurations obtained by centre transformations differ. Nevertheless, they are admissible configurations and can be chosen to form a subset. This idea has been tested various times, e.g.\ in the three-state Potts model \cite{Alford:2001ug,Kim:2005ck}
and in QCD with static quarks \cite{Aarts:2001dz}. Recently a detailed analysis was given for QCD in one dimension and it was shown that the sign problem can be eliminated completely for five flavours or less  \cite{Bloch:2013ara}. 
 A slightly different subset construction eliminates the sign problem in random matrix theory, an effective theory for QCD at low temperature and in the Silver Blaze region before onset \cite{Bloch:2011jx,Bloch:2012bh}.

A second manifestation of this idea has been developed for theories with fermions, interacting either via a four-fermion interaction or via the exchange of  a bosonic field \cite{Chandrasekharan:2009wc,Chandrasekharan:2011mn,Chandrasekharan:2012va}.
 Here the idea is to avoid integrating out the fermions in one sweep, yielding the determinant, but instead take a more measured approach and integrate out the fermions in such a way that obvious fermion bags, with a positive weight, can be identified.  In absence of a unique way to identify those fermion bags, guidance can come from strong and weak coupling expansions. One model to which this idea has been applied successfully, is the massless Thirring model in three dimensions \cite{Chandrasekharan:2011mn}. While ordinary Monte Carlo algorithms have great difficulty in approaching the chiral limit, the fermion bag approach works particularly well in this case. What is lacking so far is the inclusion of nonabelian interactions. More details can be found in the recent review \cite{Chandrasekharan:2013rpa}.

\section{Change in integration order/representation}

The lesson from the previous section, especially the fermion bags, is that the appearance of the sign problem depends on the way in which the path integral is evaluated, i.e.\ 
the order of  integration. This is in fact more generally true: the sign problem can be severe in one representation of the path integral but manifestly absent in another. Again this idea is not new, it was already seen in the strong-coupling limit of QCD, where the gauge links are integrated out first and the remaining fermionic theory is written as a sum over gauge-invariant monomers, dimers and closed baryon loops \cite{Karsch:1988zx}. The fermion determinant never appears and the sign problem is absent, or at least much milder than before.
This observation has been successfully combined with worm-type algorithms \cite{Prokof'ev:2001zz},  allowing for a more efficient sampling \cite{deForcrand:2009dh}.

For bosonic theories, this idea has been developed further, by performing 
strong-coup\-ling expansions to all orders, under the name of dual formulations \cite{arXiv:1104.2503}. Here the original field variables are interchanged for worldline and flux variables, with manifestly real and positive weights. This has been carried out  for a number of models, including abelian gauge theories \cite{Mercado:2013ola}. The main obstacle is again the inclusion of nonabelian interactions, see Ref.~\cite{cglat} for a recent review.

Finally, this notion is also relevant for effective three-dimensional models, constructed using combined strong coupling and hopping parameter (heavy quark) expansions, which can subsequently be studied using flux representations \cite{arXiv:1104.2503,Fromm:2011qi,Mercado:2012ue}, complex Langevin dynamics \cite{KW,Aarts:2011zn,Fromm:2012eb}, or even ordinary Monte Carlo simulations combined with reweighting, since the sign problem is milder than in the original formulation \cite{Fromm:2011qi}.

\section{Outlook}

So far the QCD phase diagram has not been determined nonperturbatively using lattice QCD, due to the sign problem at finite density.
In an attempt to solve this longstanding problem, a variety of ideas is being pursued, following logically independent starting points. Here I discussed a selection of those, based on the ideas of constrained simulations, holomorphicity, partial summations and changes in representation. 
Many of the techniques are still in development for QCD and hence are being tested in theories that are simpler than QCD, but nevertheless suffer from a sign problem. Establishing various (competing) approaches in parallel is very stimulating, since trust in new methods is greater when existing results can be reproduced. Due to the sign problem,  there is often no outstanding benchmark result:  a consistency between results obtained with new methods is then the best one can hope for.

Returning to QCD, we note that heavy quarks can be included via a hopping parameter or $1/m$ expansion. Here first results for the phase diagram are starting to appear, employing e.g.\ the histogram method \cite{Ejiri:2013lia} and the strong-coupling expansion \cite{Fromm:2012eb}. A comparison between results obtained with such different approaches would help in gaining confidence in those.
QCD with light quarks is considerably harder: here first results have been obtained with complex Langevin dynamics
\cite{Sexty:2013ica}.

\acknowledgments

It is a pleasure to thank my collaborators, especially Erhard Seiler, D\'enes Sexty, Nucu Stamatescu and Lorenzo Bongiovanni. 
I also thank Kim Splittorff for a careful reading of the manuscript, and Sourendu Gupta for the opportunity to present this overview and kind hospitality at TIFR, Mumbai. 
This work is supported by STFC, the Royal Society, the Wolfson Trust and the Leverhulme Trust.

\bibliographystyle{pramana}
\bibliography{references}

\end{document}